\RequirePackage{fix-cm}
\documentclass[smallextended]{svjour3}       % onecolumn (second format)
\smartqed  % flush right qed marks, e.g. at end of proof
\usepackage{graphicx}
\usepackage{caption}
\usepackage{rotating}
\usepackage[numbers]{natbib}
\bibpunct{[}{]}{;}{n}{,}{,}

 \journalname{Experimental Astronomy}

\begin{document}

\title{Effects of capillary reflection in the performance of the collimator of the 
Large Area Detector on board LOFT.\thanks{On behalf of LOFT collaboration}}

\titlerunning{Effects of capillary reflection in the LOFT-LAD performance}   % if too long for running head

\author{Teresa Mineo
\and George W. Fraser
\and Adrian Martindale
\and Charly Feldman
\and Riccardo Campana
\and Giancarlo Cusumano
\and Marco Feroci
}


\institute{T.~Mineo, G.~Cusumano \at
              INAF, IASF-Palermo, via U. La Malfa 153, I-90146 Palermo, Italy
             Tel.: +39-091-6809478\\
              Fax: +39-091-6882258\\
              \email{mineo@iasf-palermo.inaf.it}     
\and R.~Campana, \at
INAF/IASF-Bologna, Via Gobetti 101, I-40129 Bologna, Italy
\at
INFN/Sezione di Bologna, Viale Berti Pichat 6, I-40127 Bologna, Italy
\and M.~Feroci \at
  INAF, IAPS, via del Fosso del Cavaliere 100, I-00113 Roma, Italy
\at
INFN/Sezione di Roma 2, viale della Ricerca Scientifica 1, 00133 Roma, Italy
\and  G.W.~Fraser, A.~Martindale, C.~Feldman \at
Space Research Centre, Michael Atiyah Building, Department of Physics and Astronomy, University of Leicester,
LE1 7RH, Leicester, UK
}

\date{Received: date / Accepted: date  (Version 3.0 - 11 February 2013)}

\maketitle

\begin{abstract}
The Large Observatory For X-ray Timing (LOFT) is one of the candidate missions selected 
by the European Space Agency for an initial assessment phase in the Cosmic Vision programme.
It is proposed for the M3 launch slot and has broad scientific goals related to fast 
timing of astrophysical X-ray sources.
LOFT will carry the Large Area Detector (LAD), as one of the two core science instruments,
necessary to achieve the challenging objectives of the project.
LAD is a collimated detector working in the energy range 2-50 keV 
with an effective area of approximately 10 m$^2$ at 8 keV.
\\
The instrument comprises an array of modules located on deployable panels. 
Lead-glass microchannel plate (MCP) collimators are located in front of the large-area 
Silicon Drift Detectors (SDD) to reduce the background contamination from off-axis 
resolved point sources and from the diffuse X-ray background.
The  inner walls of the microchannel plate pores reflect grazing incidence 
X-ray photons with a probability that depends on energy. 
In this paper, we present a study performed with an ad-hoc simulator
of the effects of this capillary reflectivity on the overall instrument performance.
The reflectivity is derived from a limited set of laboratory measurements, used to constrain the model. 
The measurements were taken using a prototype collimator whose thickness is similar to that
adopted in the current baseline design proposed for the LAD.
\\
We find that the experimentally measured level of reflectivity of the pore inner walls 
enhances the off-axis transmission at low energies, producing
an almost flat-top response. The resulting background increase due to the diffuse cosmic 
X-ray emission and  sources within the field of view does not degrade the instrument sensitivity. 

\keywords{X-ray optics, X-ray collimator, microchannel plate, capillary reflection, LOFT }
\PACS{ 07.85.Fv  \and 07.87.$+$v  \and 95.55.Ka}
\end{abstract}

\section{Introduction}
\label{intro}
 LOFT \citep{feroci2010} is devoted to study  matter in extreme states such as those found at the event 
horizon of black holes or the surface of neutron stars. These sources exhibit rapid luminosity changes 
and spectral variability in the X-ray region, providing the key scientific justification for the Large Area  
Detector \cite[LAD;][]{zane2012} one of the two core scientific instruments proposed for LOFT.         

\begin{figure}[ht]
\begin{minipage}[t]{0.45\linewidth}
%\centering
\includegraphics[ width=\textwidth]{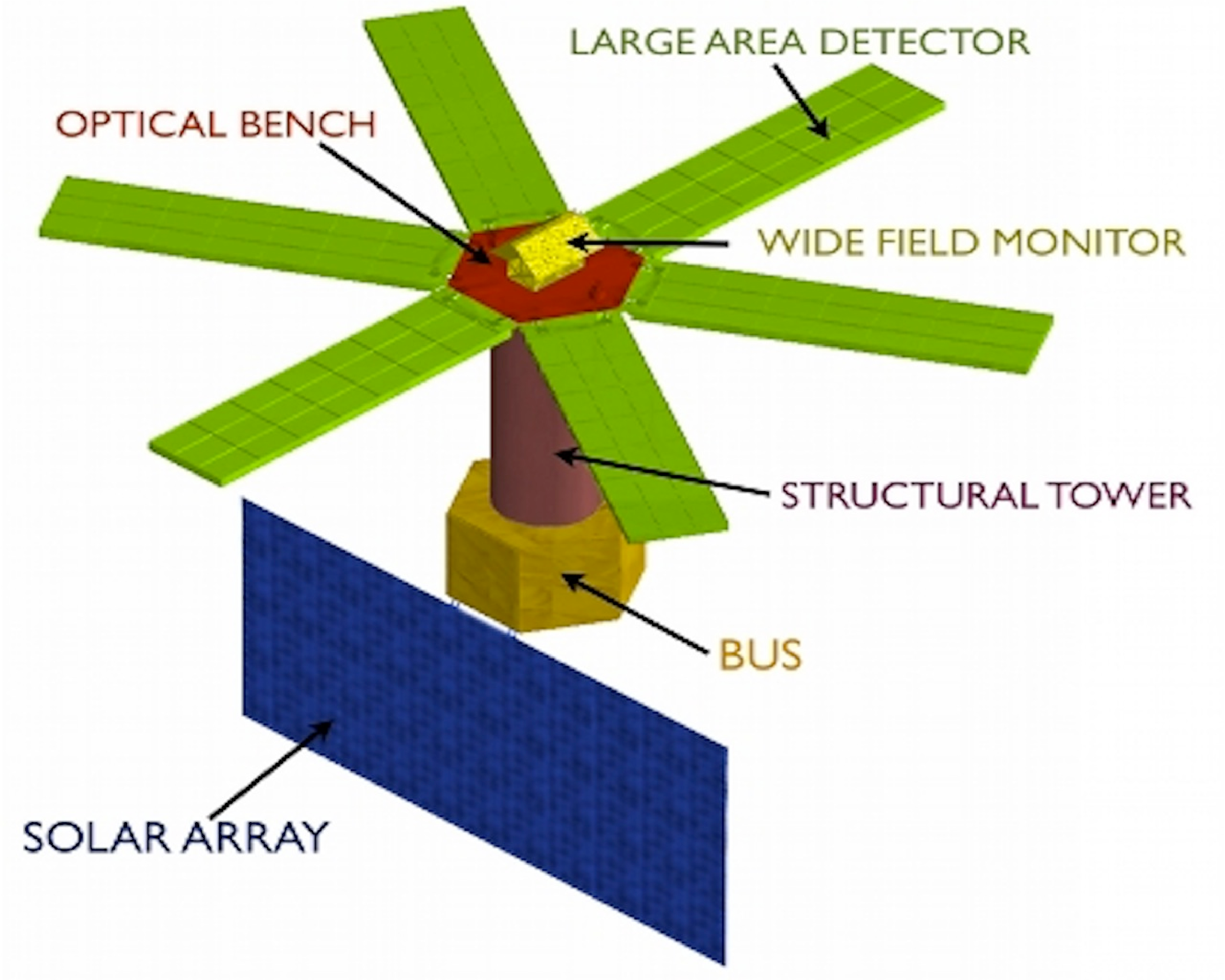}
\caption{Baseline configuration for the LOFT satellite; the six green panels together form the LAD.}
\label{satellite}
\end{minipage}
\hspace{0.5cm}
\begin{minipage}[t]{0.45\linewidth}
\includegraphics[width=\textwidth]{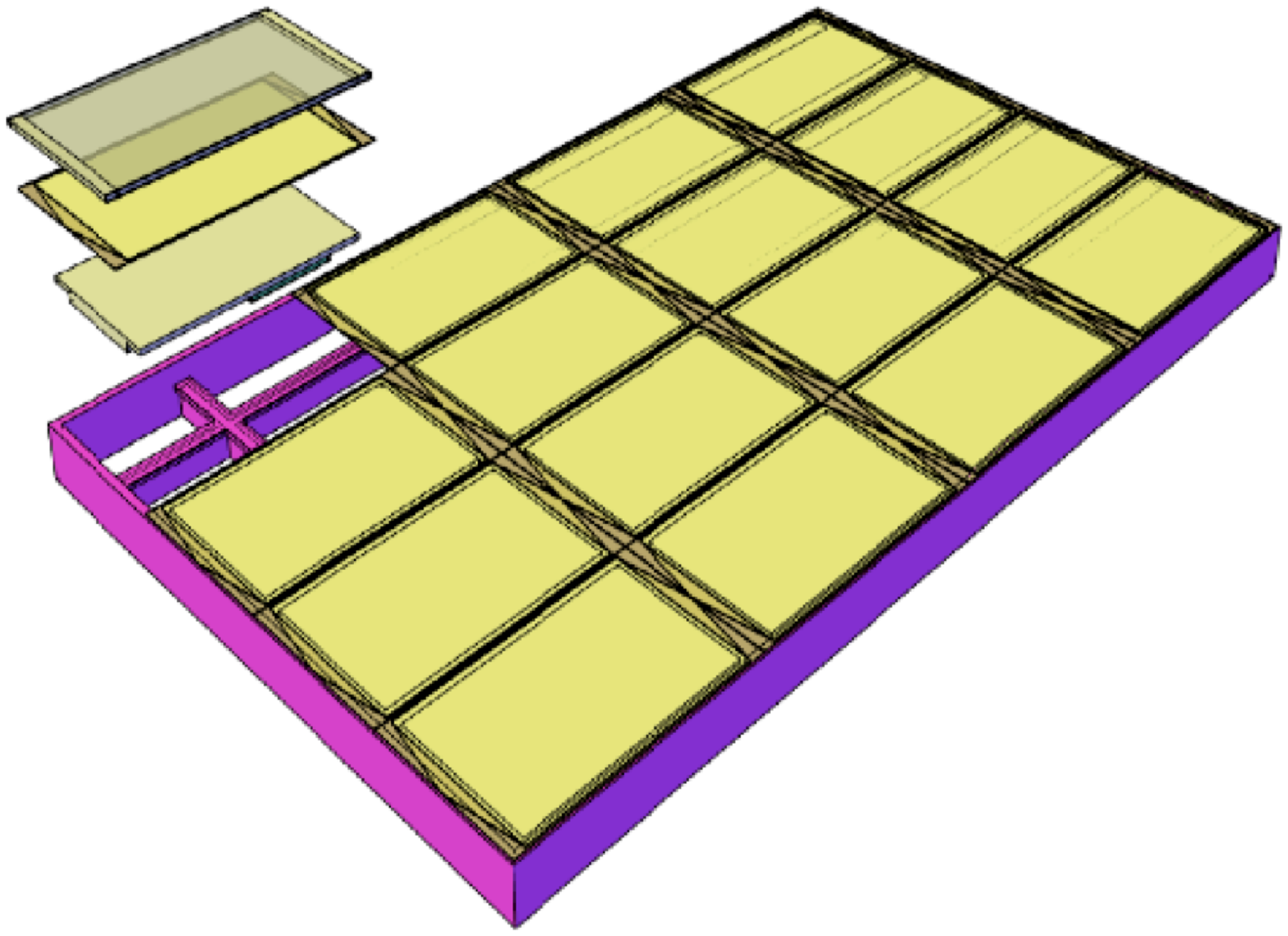}
\caption{Schematic of the LAD Module, the exploded section shows (from top down) 
the mounting of the collimator, SDD and the Front End Electronics board.}
\label{module}
\end{minipage}
\end{figure}

The LAD is a collimated detector based on Silicon Drift Detectors (SDD)
that work in the energy range 2-50 keV with an effective area 
of approximately 10 m$^2$ at 8 keV, an energy resolution  of 4\% at 6 keV and 
a timing accuracy of 10 $\mu$s.
\\
The LAD detector consists of 2016 SDDs located in six deployable
panels (see Fig.~\ref{satellite}) and organized as shown in Fig.~\ref{module} in 
modules of 16 elements mechanically and electrically linked to form a single
coherent instrument.
The SDD version adopted for LAD is an optimization of the model used in 
ALICE, an experiment of the Large Hadron Collider at CERN 
\cite{vacchi91,rashevsky02,beole07}.

The  Field of View (FoV) is limited by a collimator  placed in front of the detector, 
in order to reduce the background contamination from off-axis point sources and the 
diffuse cosmic X-ray background (CXB). 
Its design requires a trade-off as introducing a collimator drastically reduces background 
from off-axis sources, but produces a varying instrument response with off-axis angle. 
Hence, care is required to avoid spurious modulations in the detected signals caused by the 
limitations of the precision of spacecraft attitude control and by pointing stability
($\sim$1$'$ for the LAD).
\\
Following the heritage of the EXOSAT instruments MEDA \cite{turner81} and  GSPC  
\cite{peacock81} and, more recently, the collimator on the BepiColombo Mercury 
Imaging X-ray Spectrometer \cite[MIXS-C;][]{fraser10}, 
the technology adopted for the LAD is based on lead-glass microchannel plates (MCP). 
This type of collimator  provides a transmission function that is suitable for
the scientific objectives of LOFT mission, while  minimizing 
the mass per unit area - as required by the large surface area that most be covered for LOFT  
(active area $\sim$15 m$^2$).

An effect of primary importance for LAD performances is that X-ray photons with small 
grazing incident angles (lower than a few degrees)  can be reflected by the inner walls of the pores in 
lead glass collimators according to the total reflection theory \cite{willingale98, fraser11, kaaret92}. 
This effect increases  the collimator transmission at off-axis angles, rounding or flattening the 
transmission profile,  but also enhancing the background contribution from CXB and field sources - 
degrading instrument sensitivity.

In this paper, we present an ad-hoc study to evaluate the influence of channel reflectivity on the 
instrument performance. Section \ref{collimator} presents details of the collimator chosen in the 
LAD baseline design. Section \ref{sect-refl} presents a limited experimental measurement of a 
prototype collimator, used to constrain the reflectivity model reported in Section \ref{code}. 
Section \ref{sims} presents the results of a simulator for the instrument's response 
(including the transmission function of the collimator and the level of detected CXB) and Section 
\ref{summary} describes the conclusions of our study.

\section{The collimator}
\label{collimator}
The collimator adopted for the LAD is summarized in 
Table~\ref{tabcoll} and comprises an array of 8$\times$11 cm MCPs which are
held $\sim$2 mm above the SDD input face by a Titanium frame.  Sixteen SDDs and 
collimators are held in an Al detector housing which constitutes a module (Fig.~\ref{module}). 

MCPs are made by a draw and etch process similar to how optical fibres are produced \citep{martin00}. 
A lead-glass core, is surrounded by a cladding glass, forming a couple which is drawn and 
stacked to produce a multi-fibre. These are then stacked and fused to produce a block, which 
which can be sliced, polished to the required thickness and etched to produce the finished plates.
The wall of the pores are therefore made of the cladding glass, with a density of 3.3 gm cm$^{-3}$ 
and Pb fraction ($\sim$37\% by weight)  guaranteeing good absorption of X-rays in the pore septa 
in the 2-30 keV energy band of LAD. 

This configuration gives an open area ratio of  $\sim$70\%, an aspect ratio of 60:1 and a 
geometrical FoV of $\sim$1$^\circ$.  The transmission function presents a moderate 
dependence on the azimuthal angle $\phi$ (relative to the axes defined by the Cartesian pore 
walls) because of the square shape of the pore. 
In particular, moving from $\phi$=0$^\circ$ to $\phi$=45$^\circ$,  the geometrical collimator 
transmission is reduced by $\sim$4\% at 5$'$  and by $\sim$15\% at 15$'$.

 A blanket of 80 nm Al supported by 1 $\mu$m Kapton, placed
either above or below the MCPs,  shields LAD against UV, optical, and infrared 
light contamination. Moreover a second Al film, 80 nm thick, is deposited on top
to thermally insulate the collimator and provide a second layer of light rejection. 
This second layer offers added resilience against light ingress through pinholes in the foils.

The collimator transmission function is  influenced by misalignments coming 
from a number of sources as mounting tolerances  between the optical axis and the 
deployable panels, or  between the separate collimators or modules. The pore-to-pore 
alignment in the MCP itself is not negligible. Moreover, time variable effects exist, such as those 
induced by  thermal deformations of panels and modules during the different thermal loading 
around an orbit.
\\
To maintain LAD performances within the requirements derived from the
proposed scientific objectives, the  total misalignment error budget must be of order 
2.5$'$ where the contribution from pore orthogonality is $\sim$1$'$.

The polishing of the couple prior to drawing and the drawing process itself lead to pore surfaces 
that are intrinsically smooth. Subsequent removal of the core glass by acid etch \cite{wallace06,wiza79} 
produces a silicon-rich layer in the inner surface of the pores where lead and other elements have been 
leached out  \cite{dsuza96}. Moreover, a roughening of the inner pore surface may occur, whose level 
depends on several factors, including the etching uniformity, irregularity of the core glass surface or 
diffusion effects across the glass boundaries \cite{kaaret92}. These combined effects lead to the reflectivity 
of the channel pore wall and must be accounted for in the theoretical model described below.

\begin{table}
\begin{center}
\caption{Characteristics of the LAD collimator}
\begin{tabular}{ll}
\hline\noalign{\smallskip}
Parameter& Value  \\
\noalign{\smallskip}\hline\noalign{\smallskip}
pore size  &	100.0  	$\mu$m   \\
septal thickness  &	20.0  $\mu$m \\
collimator thickness  &	    6.0   mm \\
open area fraction & 70\% \\
aspect ratio & 60:1 \\
Al film thickness & 80.0  nm \\
blanket Kapton thickness	  & 1.0   	$\mu$m \\
blanket Al thickness	 & 80.0  nm \\
collimator-focal plane distance	& 2.0	mm \\
total misalignment error budget & 2.5$'$\\
\noalign{\smallskip}\hline
\end{tabular}
\label{tabcoll}      
\end{center}
\end{table}

\section{Laboratory measurements }
\label{sect-refl}
\begin{figure}
\begin{minipage}[t]{0.49\textwidth}
\centering
\vspace{-2 cm}
\includegraphics[angle=-90,width=0.89\textwidth]{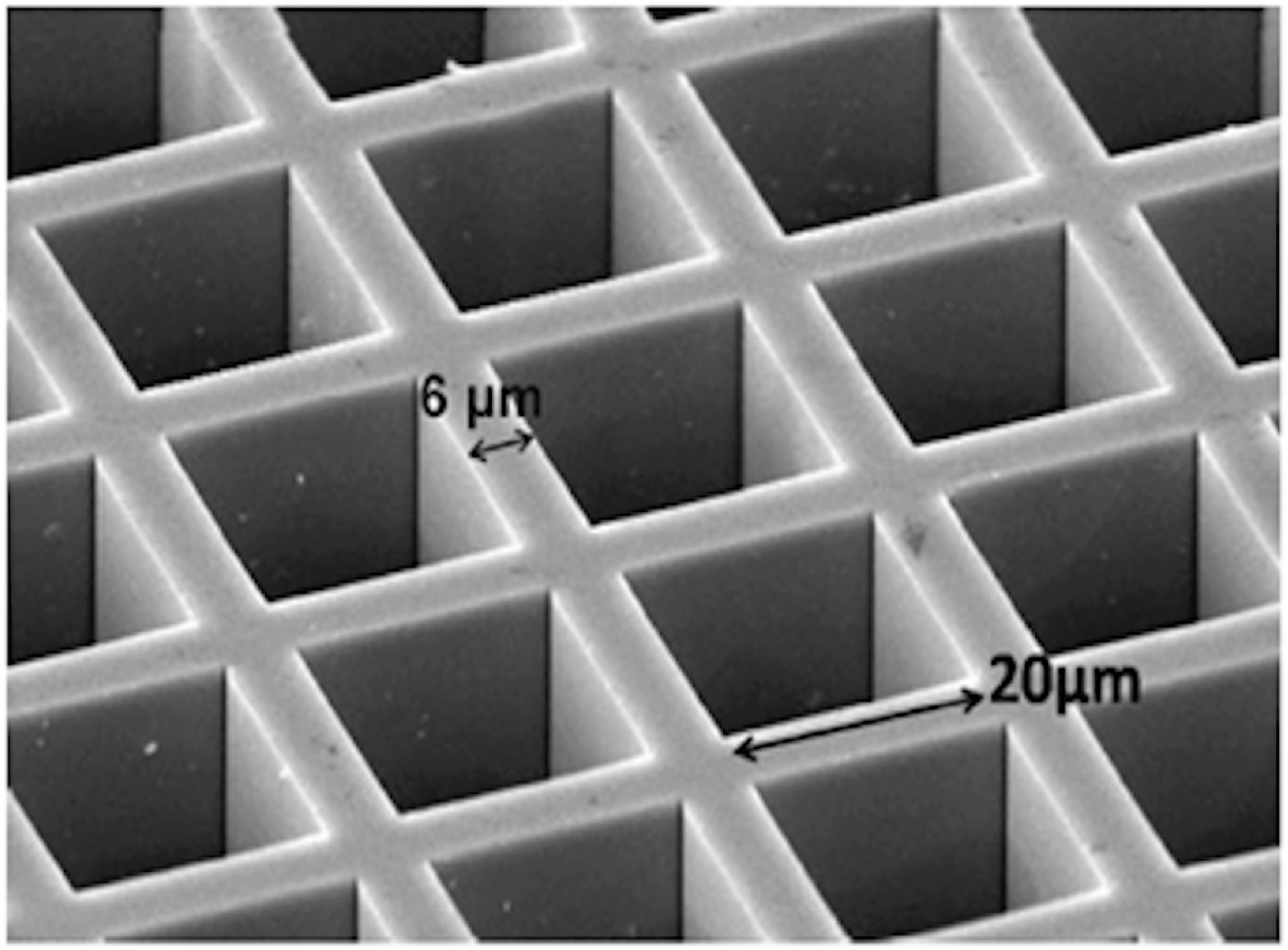}
\captionof{figure}{Electron microscope image of the collimator 
used in the laboratory measurements for evaluating pore reflectivity.}
\label{ima_coll}
\end{minipage}
\parbox[!t]{0.49\linewidth}{
\centering
\captionof{table}{Relative transmission 
at 2.4 KeV from laboratory measurements}
\begin{tabular}{cc}
\hline
\multicolumn{1}{c}{Angle } & \multicolumn{1}{c}{Transmission} \\
\multicolumn{1}{c}{(arcmin) } &  \\
\hline
 ~0     &   1.00$\pm$0.020  \\
 ~5     &   0.89$\pm$0.018 \\   
10    &  0.82$\pm$0.017 \\
15    &  0.66$\pm$0.015 \\
20    &  0.42$\pm$0.011 \\
25    &  0.17$\pm$0.006 \\
30    &  0.07$\pm$0.004 \\
35    &  0.03$\pm$0.002 \\
\hline
\end{tabular}
\label{lab_data}   
}
\end{figure}

We have evaluated the capillary reflectivity of a representative sample collimator which was
5mm in thickness - similar to the final design value for the LAD collimator.
This sample has a 250:1 aspect ratio, 20 $\mu$m pore width and 6 $\mu$m septal thickness. 
No Al film is deposited on its input surface (as called for in the LAD design), 
however, this should have no influence on the comparison to the Monte Carlo model which is of primary 
importance here. An electron  microscope image of the sample is shown in Fig.~\ref{ima_coll}.  

The measurements were performed in the University of Leicester's $28$m long beamline. 
At the X-ray source end, an electron bombardment source with a Molybdenum anode was used 
to stimulate the Mo-L fluorescence lines with an average energy of $\sim$2.4 keV. 
An accelerating potential of 4keV and a 1 $\mu$m Ag filter were used to isolate the region of the emission 
lines and maximise the line to continuum ratio. 
\\
The collimator was held in a two axis rotation stage, a distance of  27 m from the X-ray source 
in order to minimise the effect of beam divergence. A large area MCP detector \cite{Lees97} 
was placed behind the collimator to measure changes in the intensity of the radiation passing through
 the collimator as a function of off-axis angle. The measurements were performed by rotating the 
collimator incrementally by steps of 5$'$. 
The ratio between the detector count rate at each angle and the on-axis value are shown in  
Table~\ref{lab_data}. 

The errors in the count rate are dominated by the dark noise of the detector,  where radioactive 
isotopes in the detector glass lead to a non-zero background. The errors reported in the table reflect this, 
of paramount importance here is the ratio of the input count rate and the detector noise over the relevant 
area of the detector. Although the source generates a large emission rate for X-rays, the solid angle 
subtended by the collimator at 27 m is small, making it difficult to improve the signal to noise 
significantly without compromising on beam divergence. 

\section{The code}
\label{code}
\begin{figure}
\centering
\includegraphics[angle=-90,width=14.0cm]{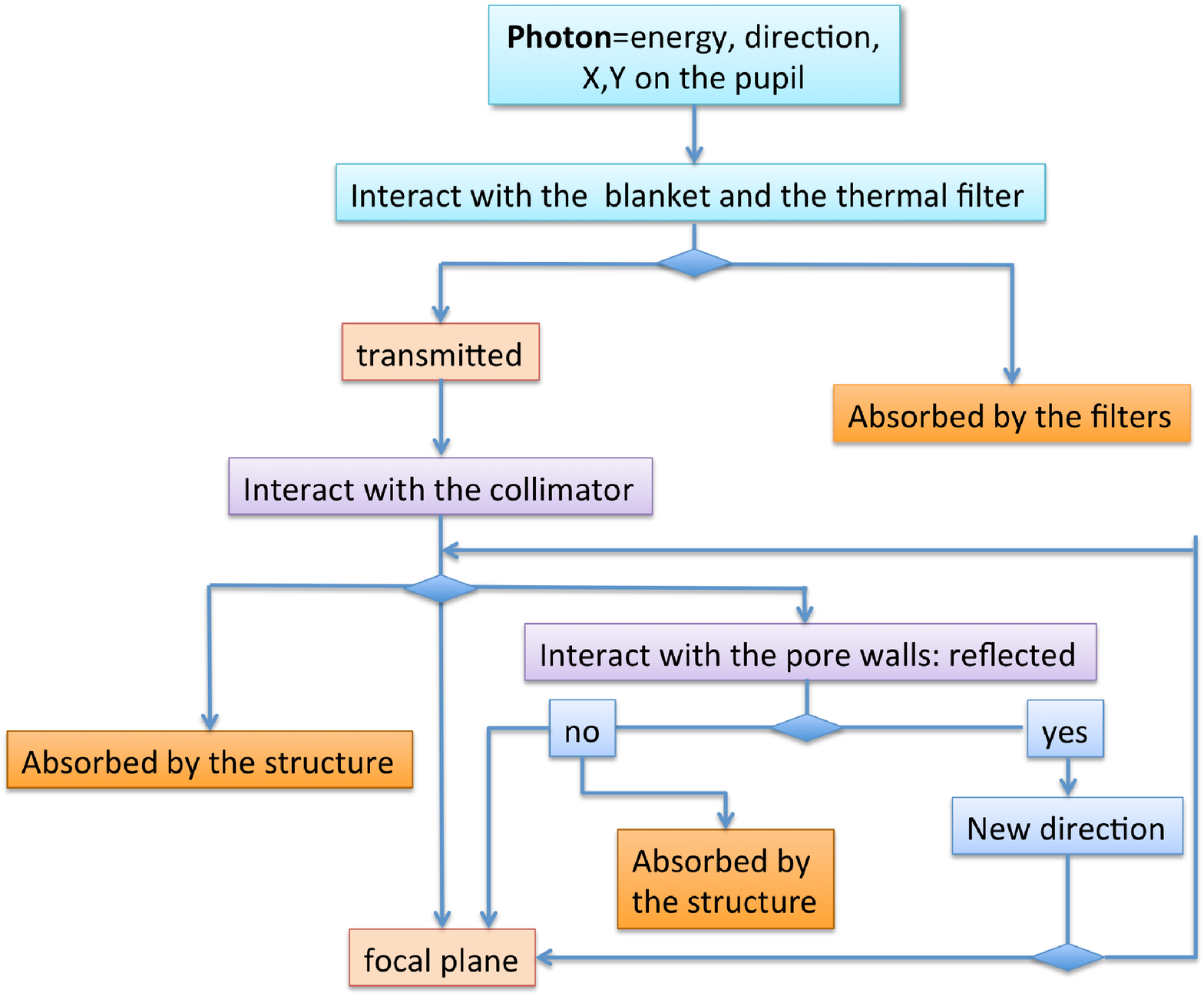}
\caption{Ray-tracing flow chart.}
\label{raytr} 
\end{figure}

The experimentally obtained transmission function includes several effects which 
cannot be deconvolved easily during data analysis, including: beam divergence, MCP pore 
misalignment and pore wall reflectivity.  
 A ray-tracing simulator allows a mathematical approach to deducing the major contributors 
to the transmission function, by computing a transmission function for comparison with the 
empirical data. 
\\
The simulator is a stand-alone ray-tracing code which models the interactions of X-ray photons 
with the collimator and follows them to the detector surface. The photon interaction processes, 
in all stages, are governed by  Montecarlo calculations.
The flow chart of the code is presented in Fig.~\ref{raytr}. 

The collimator is defined as a solid parallelogram with a porous structure,  where each pore is 
described as an empty parallelogram identified by the coordinates of its centre and by the 
width of the sides.  Both filters are simulated directly in front of the collimator. 
MCP pore misalignments are  taken into account by varying the photon incident direction by 
an angle randomized from a Gaussian distribution whose standard deviation 
is given as an input parameter.
\\
Source photons can either be simulated with monochromatic energy or drawn from 
spectral distributions introduced as tables, the nature of the source can be either 
point-like or diffuse and it may be located at either finite or infinite distance 
from the collimator. The modelled detection surface has the same geometrical area 
as the collimator and no detector response is included. 

Each photon is characterised by a direction, an energy and an impact position 
randomly distributed on the collimator surface. 
If not absorbed by the filters, photons can be either directly transmitted to the focal plane 
or interact with the collimator. In the latter case, they can either be absorbed by the structure 
or interact with the pore wall: multiple interactions within a pore can happen before the photon 
is absorbed or reach the detection surface. If a photon is reflected, its output direction is 
computed from Snell law.

\begin{figure}[t!]
\begin{minipage}[t]{0.49\textwidth}
\includegraphics[angle=-90, width=0.95\textwidth]{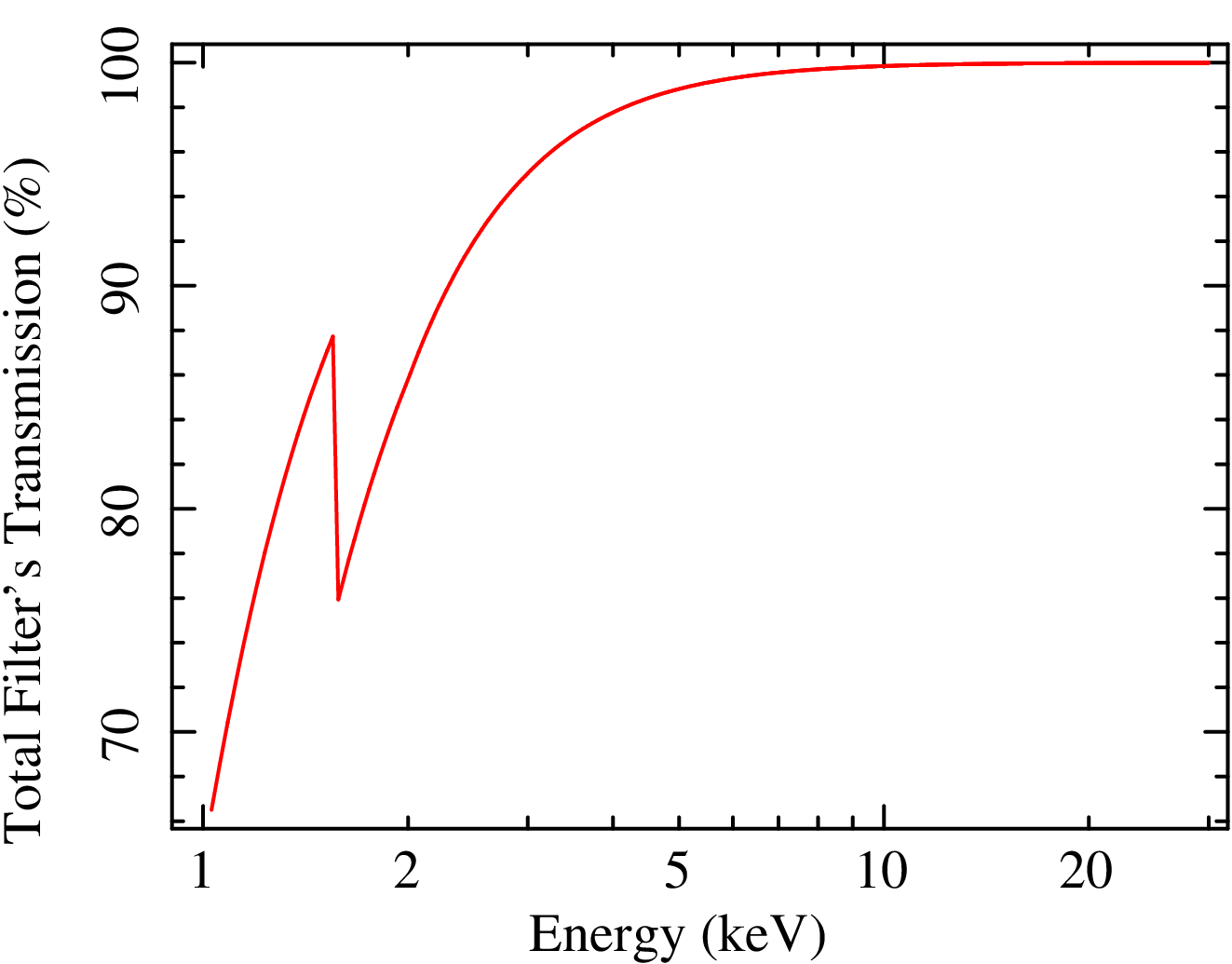}
\caption{Total transmission efficiency of the two filters applied to the collimator.}
\label{filter}
\end{minipage}
\hfill
\begin{minipage}[t]{0.49\textwidth}
\includegraphics[angle=-90, width=0.95\textwidth]{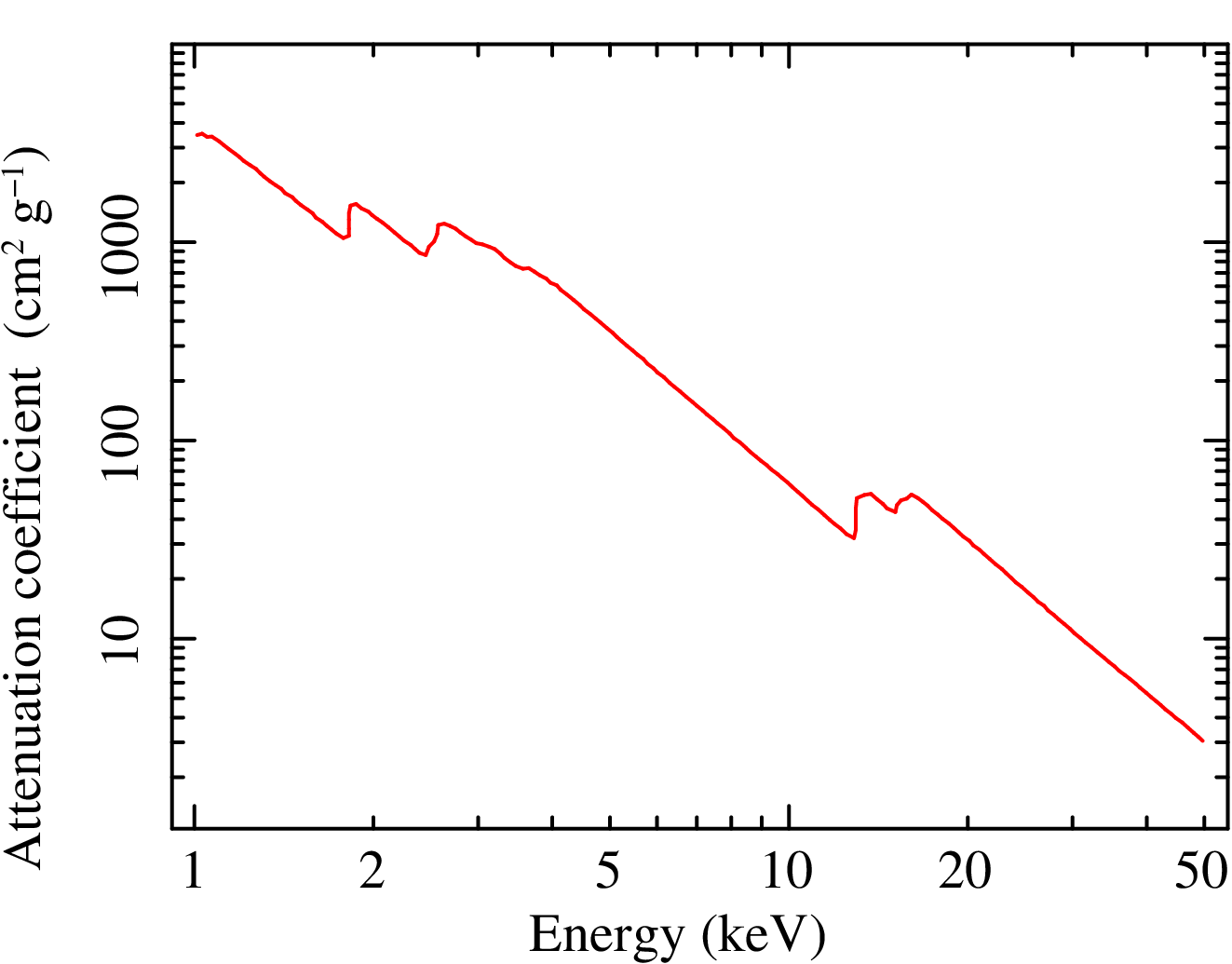}
\caption{Attenuation coefficient of the lead-glass used in the simulator.}
\label{glass_att}
\end{minipage}
\end{figure}

The  filter transmission depends on energy and incident angle and is computed 
evaluating the thickness crossed by the photon and interpolating the cross-sections 
of the involved materials from tables retrieved from the National Institute of 
Standards and Technology  (NIST) database\footnote{http://www.nist.gov/index.html}.
Figure~\ref{filter} shows the total transmission probability of the blanket and the
Al filter as function of energy:  we note that they produce a reduction of $\sim$15\% at 
2 keV and $<$1\% at 6 keV in the collimator effective area. 

In the case of the lead-glass, we used the attenuation coefficient plotted in Fig.~\ref{glass_att} 
also retrieved from the NIST database using the chemical information for the specific glass used.
The thickness of the glass crossed by the incident photon takes into account the path through 
the pore septal walls to accurately simulate transmission through the collimator structure.

\subsection{The reflectivity model}
The database of the Lawrence Berkeley National Laboratory's Center for X-Ray 
Optics\footnote{http://henke.lbl.gov/optical\_constants} provides the reflectivity 
at different values of the surface micro-roughness for several materials.
\\
We derived tables for different levels of surface micro-roughness from this database 
in order to allow us to evaluate the effect of micro-roughness on the reflection probability, 
both as function of the photon energy and incident angle. 
Tables corresponding to lead-glass were used initially to obtain the results presented in 
Campana et al. (2013).
%\citet{campana13}.
However, considering that the heavier elements are leached during the etching process
\citep{dsuza96},  we used reflectivity tables for SiO$_2$ in the simulations presented in 
this paper. 

The analysis methodology applied in our work assumes that the level of micro-roughness is 
sufficient to characterize the reflectivity i.e. micro-roughness is dominant. 
This hypothesis is based on the observation that the fraction of photons that reach the focal 
plane after reflection is lower than 20\% at any incident angle. Moreover, only 2\% of these 
scattered photons  experience  multiple scatterings.

We computed the transmission function at each value of micro-roughness  and compared 
it to the empirical data, choosing the best fit as the one that minimises the root mean square 
($rms$) of the differences between the measured and the simulated data.   
We adopted this as our reflectivity model in the calculations that follow.  In these simulations, 
we assumed  1$'$ as level of misalignments  in the pore orthogonality \cite{kaaret92}. 

\begin{figure}[tb!]
\begin{minipage}[t]{0.49\textwidth}
\vspace{0.3 cm}
\includegraphics[angle=-90, width=0.95\textwidth]{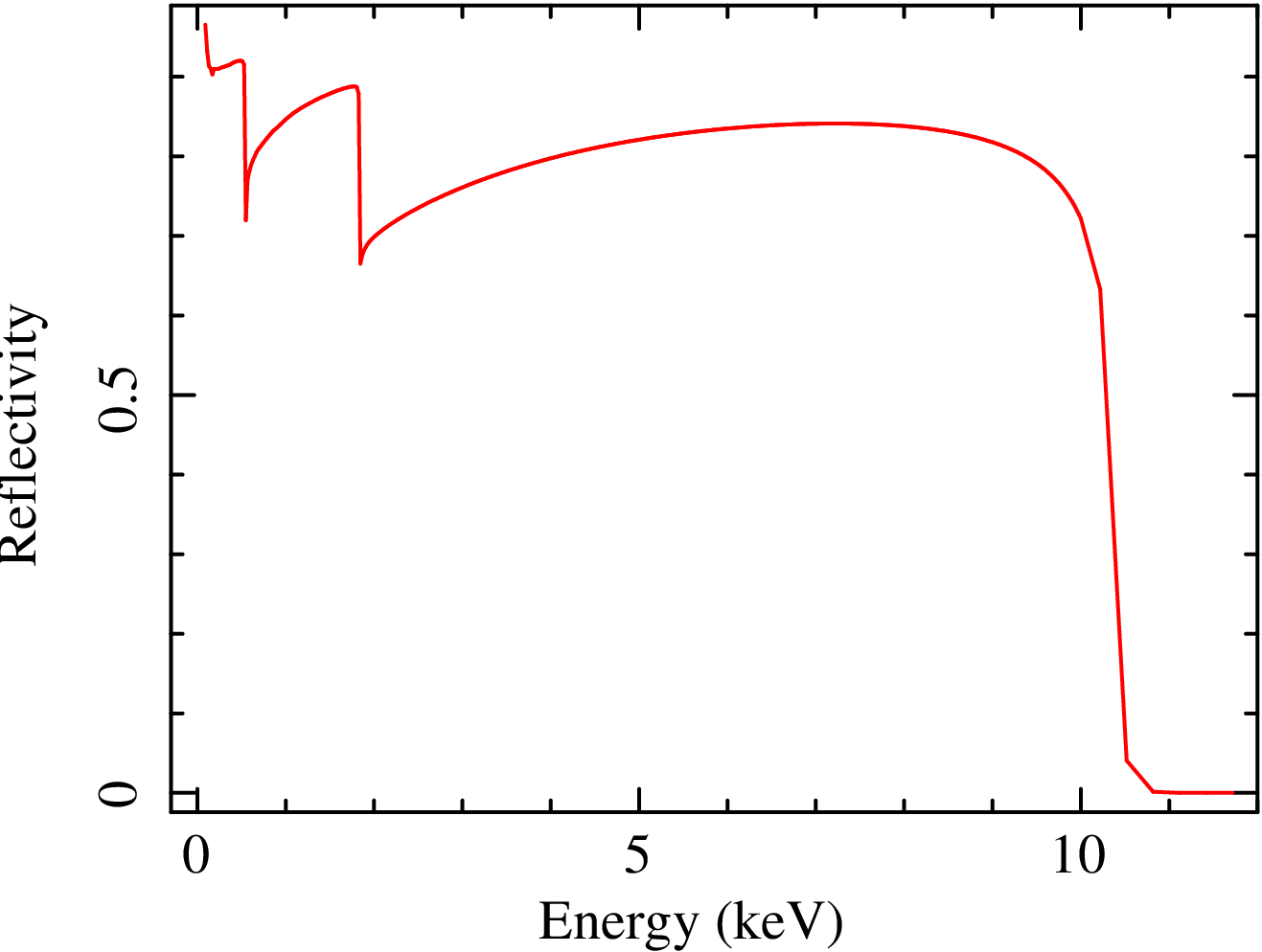}
\caption{Glass reflectivity  relative to a surface micro-roughness of 14 nm vs energy
for an incident angle of 10$'$.}
\label{reflect_ene}
\end{minipage}
\hfill
\begin{minipage}[t]{0.49\textwidth}
\vspace{0.3 cm}
\includegraphics[angle=-90, width=0.95\textwidth]{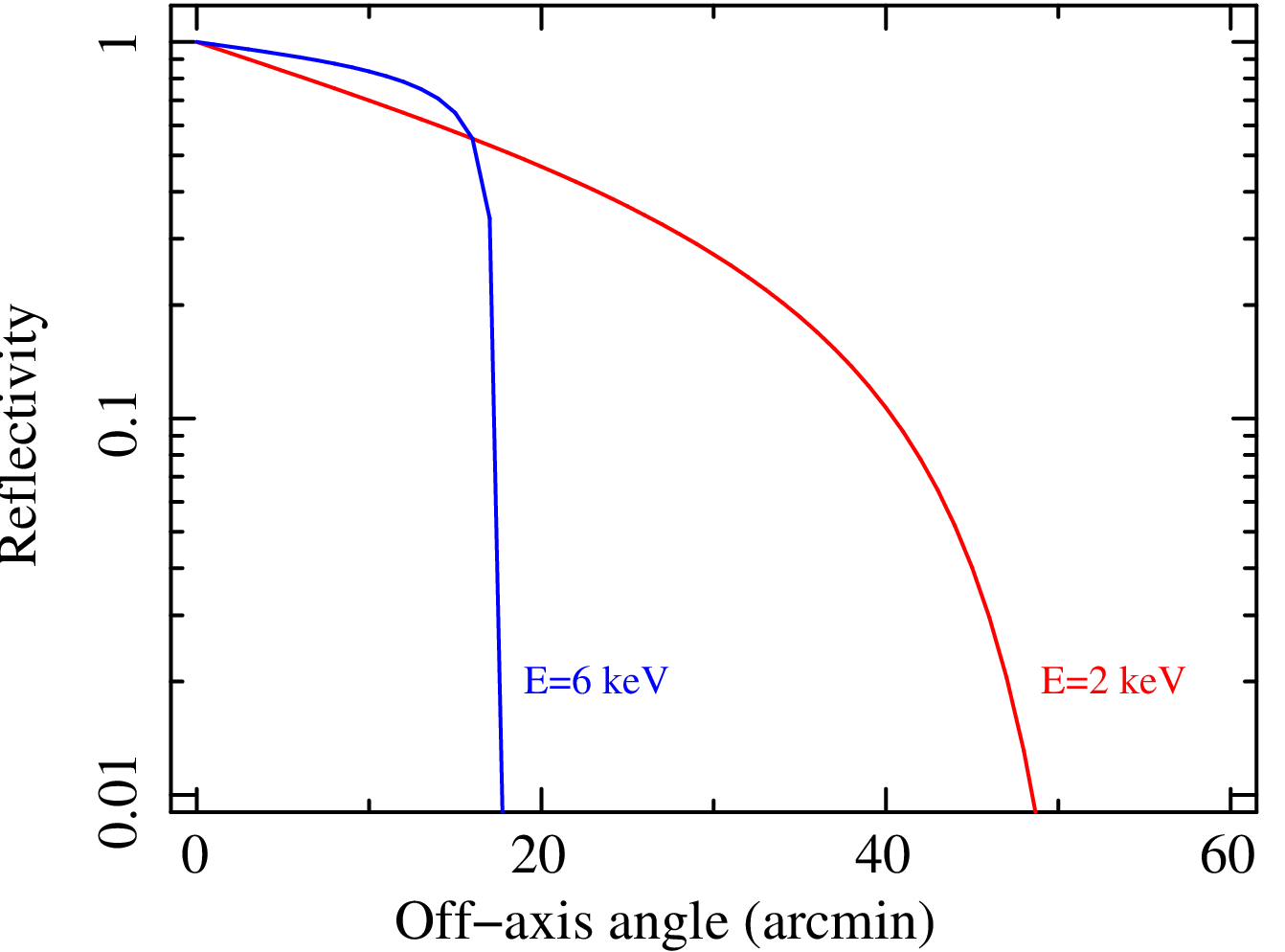}
\caption{Glass reflectivity  relative to a surface micro-roughness of 14 nm vs off-axis angles
for  2 keV (red curve) and 6 keV (blue curve).}
\label{reflect_ang}
\end{minipage}
\end{figure}

\begin{figure}[tb!]
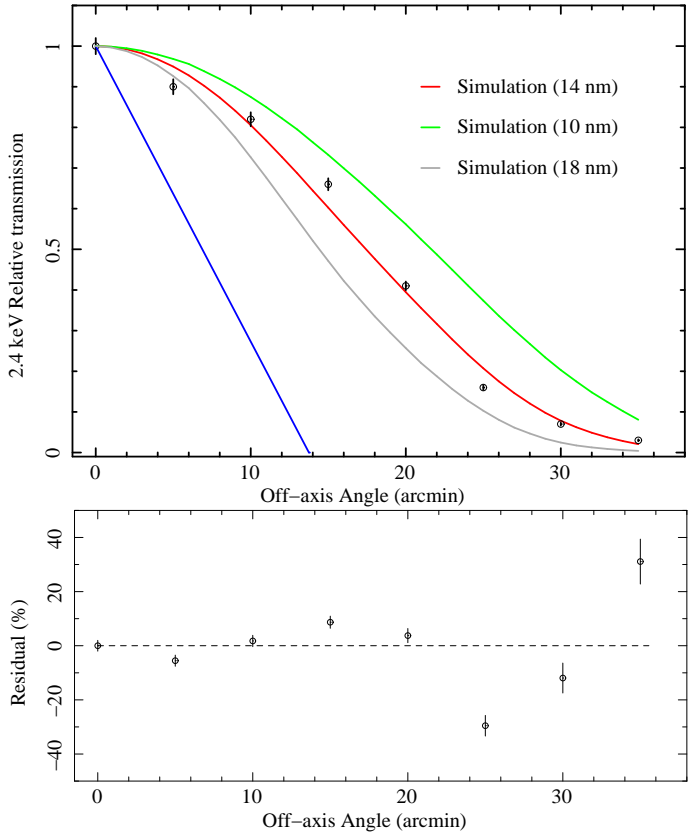

\centering
\includegraphics[angle=-90,width=0.75\textwidth]{ae_2p4_rms14_new.ps}
\includegraphics[angle=-90,width=0.75\textwidth]{ae_2p4_rms14_res.ps}
\caption{{\it Top Panel:} The red curve shows the transmission function obtained with the simulator 
fixing the micro-roughness level to 14 nm vs the off-axis angles. Lab data  are indicated with black points.
The green and grey curves represent the transmission function obtained fixing the micro-roughness level 
to 10 nm  and 18 nm, respectively.
The triangular response expected for a 250:1 aspect ratio  collimator is also shown with the blue line.
{\it Bottom Panel:} Differences in percentage between the measured and the simulated values
normalized to the measured ones  vs the off-axis angle. }
\label{reflab_1}
\end{figure}

We found that data are best reproduced by the reflectivity profile from 14 nm surface micro-roughness.
This level of micro-roughness is, quite significantly higher than values quoted in literature for lead glass 
MCPs \cite{kaaret92, fraser92} and is probably due to the very large aspect ratio of the channels in this 
collimator.
The reflection probability is shown in Fig.~\ref {reflect_ene}, for an incident angle of 10$'$, 
as function of energy and in Fig.~\ref {reflect_ang} as function of the off-axis angle for 
2 keV and 6 keV photons. 
\\
The top panel of Fig.~\ref {reflab_1} presents a comparison between the relative transmission 
obtained by the simulator (red line) assuming 14 nm micro-roughness and the data derived 
from laboratory measurements (black circles) as function of the off-axis angles.
The percentage differences between the measured 
and the simulated data relative to the measured values  is displayed in the bottom panel of the same figure.
The discrepancy is  lower than 30\% for all angles and lower than 10\% up to 20$'$.
\\
For comparison, in the top panel Fig.~\ref {reflab_1} the transmission function obtained 
fixing the micro-roughness level to 10 nm (green line) and 18 nm (grey line) together with tha 
triangular response expected for a 250:1 aspect ratio  collimator (blue line) are also shown.

\section{Simulator Results}
\label{sims}
Both the capillary reflectivity described above and the intrinsic misalignments in the collimator axes
(both due to the MCP collimator itself and errors introduced by assembly and alignment tolerances 
of the modules and panels) affect the LAD sensitivity by altering the transmission function. 
These errors lead to reduced effective area for on-axis sources and increased background from off-axis 
point sources and the diffuse X-ray background reducing the overall sensitivity.
These effects are quantified here by including the reflectivity model constrained by lab data in
 the LAD simulator and by assuming a total misalignment error of 2.5$'$ (the total error budget).
In practise, this is all attributed to pore orthogonality in the simulator, however, this simplification 
will have no impact on the results.  

\subsection{Transmission function}
\begin{figure}[t!]
\centering
\includegraphics[angle=-90,width=0.75\textwidth]{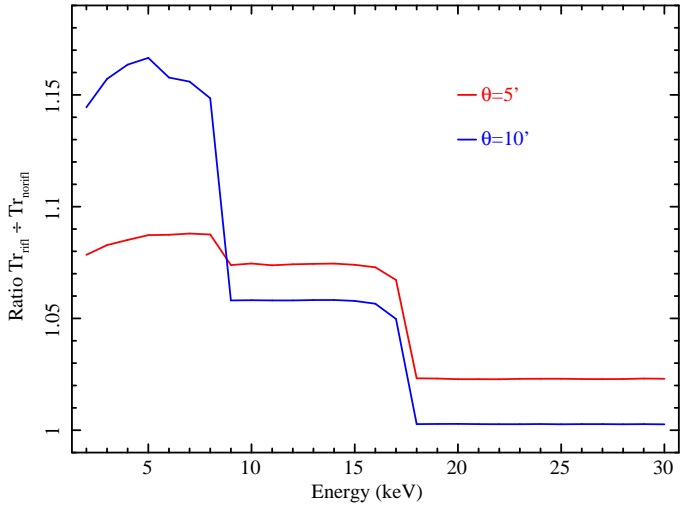}
\caption{Increasing factor of the collimator transmission
 due to the capillary reflection vs energy for a parallel beam at 5$'$  (red curve) and 10$'$
(blue curve) off-axis angles.}
\label{trasm_2}
\end{figure}

The collimator transmission function is modified by the capillary reflectivity and by misalignments.   
The misalignments introduce a reduction of the effective area on-axis and increase the amount of 
background collected from off-axis, this effect is almost independent of energy and mainly confined 
to small incident angles determined by the overall alignment errors. The effect of reflectivity is more 
subtle. For a perfect collimator with no alignment errors, it has no effect on-axis as no area of channel
 walls are projected towards the source, however, it does allow increased collection of background from 
off-axis. For a real collimator, with intrinsic misalignments, the effect is more complex - for on-axis 
sources the reflectivity allows rays to continue to reach the detector plane even though they have hit a 
wall, meaning the effect of misalignments is decreased. However, it is generally true that significant 
reflectivity will reduce sensitivity as more background will be collected.
\\
For the values adopted in the simulations, the largest reduction  in effective area due to misalignments 
($\sim$5\%) is  completely cancelled in the energy range 2 --17 keV by the reflectivity 
enhancement. This is shown in Fig.~\ref{trasm_2} where the ratio between the transmission 
with and without reflectivity is plotted as function of energy for two incident angles. 

\begin{figure}[b!]
\centering
\includegraphics[angle=-90,width=0.75\textwidth]{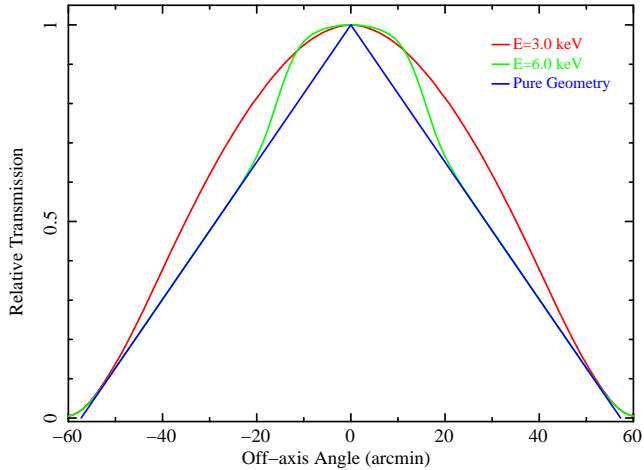}
\caption{Transmission function vs off-axis angles:
the red curve is relative to the energy of 3 keV; the green curve to the energy 6 keV. 
For comparison, the canonical transmission curve obtained without reflection  is shown in
blue. }
\label{trasm_1}
\end{figure}

The  transmission function including reflectivity and misalignment effects  is 
shown in Fig.~\ref{trasm_1} for a source at infinite distance with monochromatic 
 X-rays of 3 keV (red curve) and 6 keV (green curve).  
For comparison, the canonical triangular shape  is  also shown with a blue curve.
We note that in both curves the reduction of the transmission function with respect 
to the on axis value  is lower than for the triangular one at small angles ($<10'$).
In particular,  the decrease of the effective area  is less than 5\% up to 10$'$ for 3 keV photons.

Considering that X-ray spectra decrease with energy,  we evaluated the reduction expected in 
the observed rate of a typical source located  5$'$ off-axis whose spectrum is modelled by a 
power law with photon index  $\alpha$=2.
We found that the rate is lowered by $\sim$1\% with respect to the on axis value, and 
comparing this to the  reduction expected from a pure geometrical collimator ($\sim$8\%),
the adopted level of reflectivity  produces a flat response within the  alignment accuracy 
angles (at least at 6keV), meaning that spurious modulations due to pointing accuracy are minimised.


\subsection{Diffuse Background}

 A drawback of this flatter transmission function is the increase of the diffuse background  
that reaches the focal plane with a consequent degradation in sensitivity.
We evaluated  the effect of the capillary reflectivity on the level of the 
transmitted CXB simulating a uniformly distributed diffuse source with photon directions 
randomizes up to an off-axis angle of 70$'$ and  energies $E$ drawn in the range 2--30 keV 
according to the model  \cite{gruber99}:

\begin{equation}
\Phi(E)=7.877 \, E^{-1.29} \, \exp(-E/41.13)  \hspace{1cm} {\rm photons~cm^{-2} s^{-1} sr^{-1} keV^{-1}}
\end{equation}

\begin{figure}[t!]
\centering
\includegraphics[angle=-90,width=0.75\textwidth]{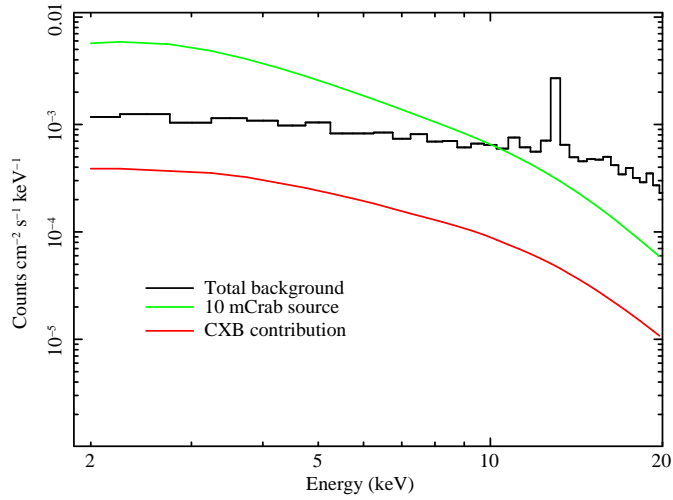}
\caption{The red line indicates the simulated spectrum of the  CXB observed by the LAD.   
The black line is the total background obtained by \cite{campana13}. For comparison, the spectrum of
 a 10 mCrab source is shown with the green line. }
\label{geant4}
\end{figure}

\noindent
The CXB rate at the focal plane is obtained by convolving the spectrum transmitted by the collimator 
with the detector response matrix: the result is shown in Fig.~\ref{geant4} (red line). 
This result agrees (within the statistical errors) with the spectrum published in Campana et al. 
(2013) %\citet{campana13}
obtained using the reflectivity of lead-glass.
\\
Comparing the spectrum with the one obtained without pore wall reflectivity and misalignments,
the highest increase is observed below 4 keV where an increase of $\sim$35\%
is detected. At higher energies, the reflectivity contribution reduces and the CXB increase is 
$\sim$20\% in the range 4--6 keV and only $\sim$10\% in the range 6--10 keV.

\begin{figure}[bt!]
\begin{minipage}[t]{0.49\linewidth}
\centering
\includegraphics[angle=-90,width=0.89\textwidth]{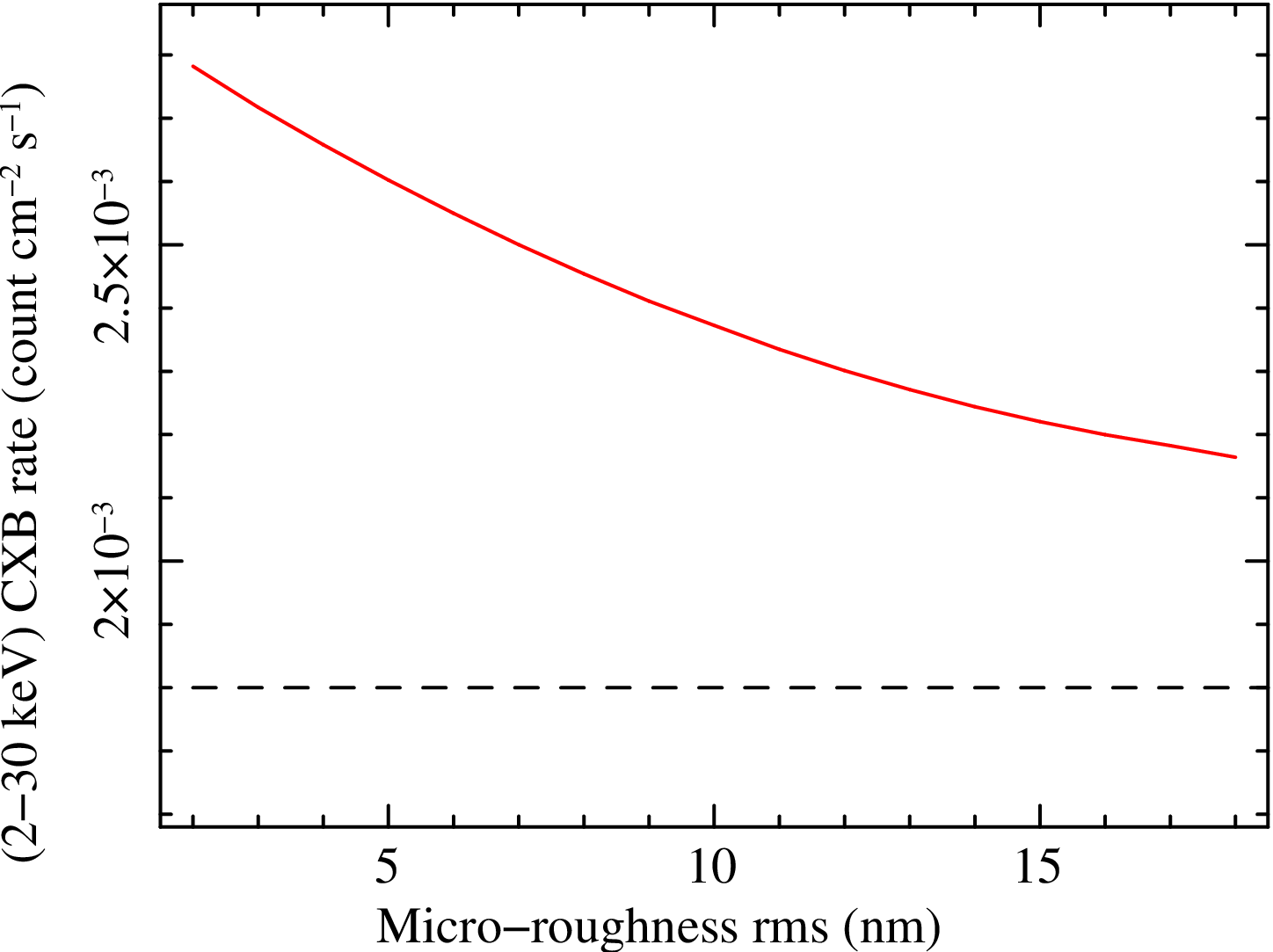}
\caption{ Rate of the CXB at the focal plane  vs the $rms$ of the surface micro-roughness.
The dashed line indicates the rate obtained without capillary reflectivity.}
\label{diffuce}
\end{minipage}
\hfill
\begin{minipage}[t]{0.49\linewidth}
\centering
\includegraphics[angle=-90,width=0.89\textwidth]{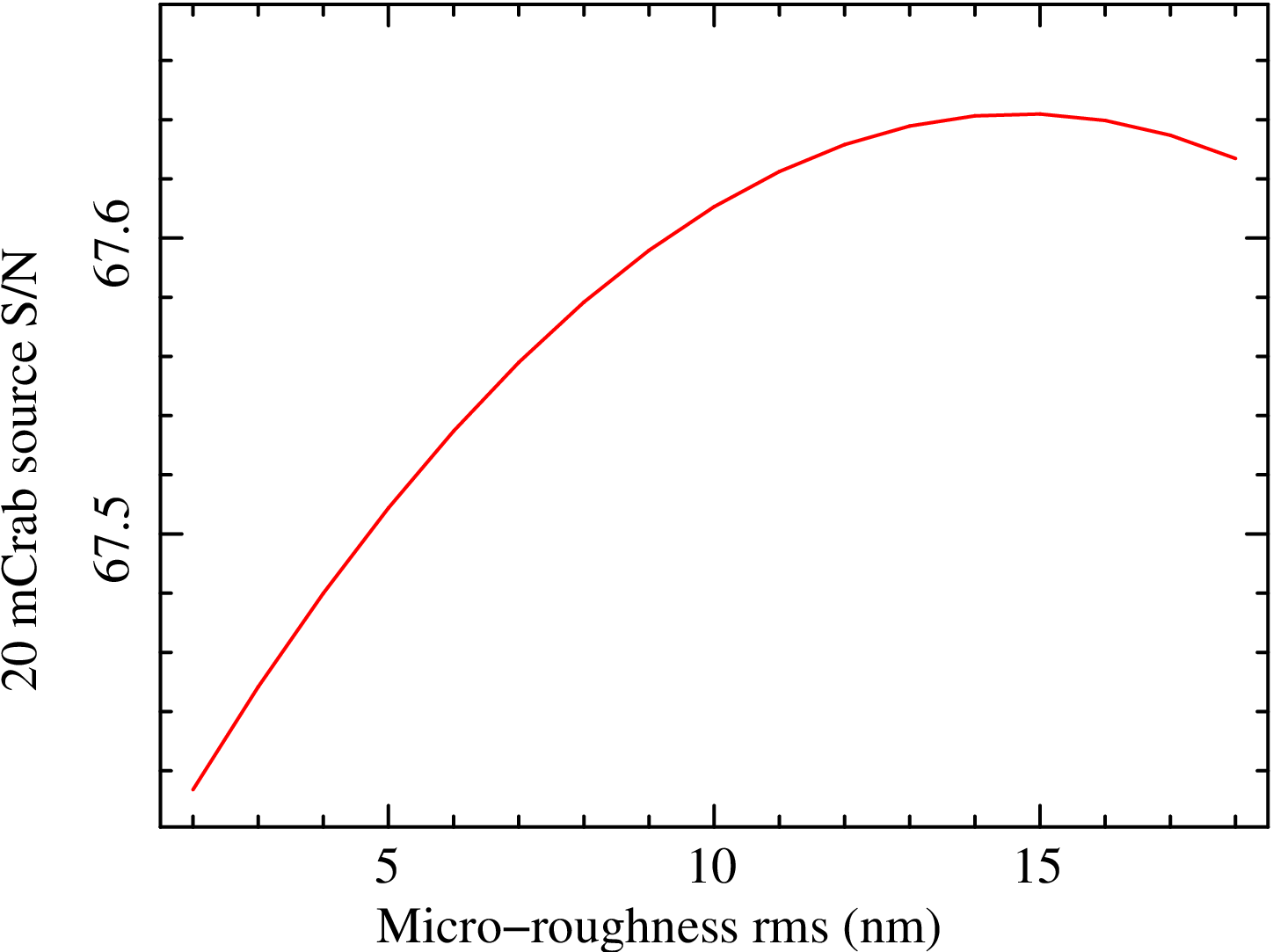}
\caption{ Signal to Noise ratio for a 20 mCrab source vs the $rms$ of the surface micro-roughness. }
\label{sig}
\end{minipage}
\end{figure}

\subsection{Discussion of the results}

We found  that reflectivity induces a significant increase of the transmission function 
at low angles. Consequently, the CXB background is increased dependant on the energy and 
direction of the incident photon.

LOFT's scientific objectives require the observations of sources up to
fluxes of   1-10 mCrab (10$^{-11}$ -- 10$^{-12}$  erg cm$^{-2}$ s$^{-1}$ ).
This  can be achieved  if the level of background is kept lower than 
10 mCrab  (equivalent to 1.6$\times$10$^{-2}$ counts cm$^{-2}$ s$^{-1}$) in the nominal 
2--30 keV energy range. 
From a detailed study, performed with a  Geant4 simulator, 
the background rate at the satellite orbit from all expected components,   
except CXB, is 1.2$\times$10$^{-2}$ counts cm$^{-2}$ s$^{-1}$ \citep{campana13}.  
\\
The rate of the diffuse cosmic X-ray background obtained with our simulations is 
2.2$\times$10$^{-3}$ counts cm$^{-2}$ s$^{-1}$, hence, the measured level of micro-roughness 
produces a  CXB level compatible with the requirements. 
The total background and the spectrum of a 10 mCrab source (assuming the 
Crab spectrum) are shown in Fig. ~\ref{geant4} with a black and a green line, respectively.

The larger transmission function at low energies could also worsen the contamination from 
off-axis sources making  the observation  of crowded fields such  as  the Galactic bulge and plane difficult 
because of source confusion.
To evaluate this effect, we used the Swift BAT 54 month catalogue \cite{cusumano10} that includes 
1256 sources and selected 65 sources whose  flux  is $>$10mCrab in the energy range 15-150 keV.
Assuming  a power law spectrum with photon index  $\alpha$=2 (Crab-like),
 we computed the LAD rate of these sources in the range 2-30 keV and the contamination from 
all sources within an angular distance of 1$^{o}$.  The number of sources whose 
spectrum will be contaminated for more than 10\%  is statistically equivalent 
to the number obtained with the triangular transmission function.  
\\ 
We can conclude that the reflectivity measured in the lab collimator  
does not degrade  the  required LAD  performances. 

 In order to determine what impact different micro-roughness would have on signal to noise in 
the LAD band Fig~\ref{diffuce} and Fig.~\ref{sig} show the calculated level of CXB background
and the detection significance of a Crab-like source (with a flux of 20mCrab) at values of micro-roughness
 close to the measured one.
We observe that the CXB rate has a maximum variation of 20\% in the investigated range
and its value at 14 nm micro-roughness is only $\sim$30\% higher than the one obtained 
without capillary reflectivity, shown in Fig~\ref{diffuce} with a dashed line. However,
because the statistical significance of a typical 20mCrab source  improves  with $rms$, 
values of micro-roughness lower than 10 nm are not recommended. 

\section{Summary and conclusion}
 \label{summary}
We obtained a model of the reflectivity from the inner walls of the pores in  
the LAD collimator and verified them against a set of empirical  measurements. 
Lab data were best fit assuming reflectivity from  SiO$_2$ with a surface 
micro-roughness of 14 nm; it is believed that the extreme channel lengths explain the 
difference between this and prior lower published values for the surface micro-roughness.

Including these results in the simulator and taking into account the  
total misalignments assumed  for the LAD, we compute the transmission function and, 
hence, the level of CXB at the focal plane
in order to evaluate the effects of reflectivity on the collimator performance. 
\\
We conclude  that capillary reflectivity with a level of micro-roughness close to 
that measured experimentally does not degrade the collimator performance significantly.

\begin{acknowledgements}
Authors acknowledge the financial support from the Italian INAF Tecno-PRIN 2009 and 
ASI/INAF contract I/021/12/0,  and from the UK Space Agency.
\end{acknowledgements}


\begin{thebibliography}{}

\bibitem{beole07}
Beol{\`e}, S.,  Alessandro, B.,  Antinori, S.,  Coli, S.,  Costa, F.,  Crescio, E.,  
Falchieri, D.,  Arteche Diaz, R.,  Di Liberto, S.,  Giraudo, G.,  Giubellino, P.,  
Masetti, G.,  Mazza, G.,  Meddi, F.,  Prino, F.,  Rashevsky, A.,  Riccati, L.,  Rivetti, A., 
Senyukov, S.,  Simonetti, L.,  Toscano, L.,  Tosello, F.,  Urciuoli, G.~M.,  
Vacchi, A.,  Wheadon, R.,
The ALICE silicon drift detectors: Production and assembly,
Nuclear Instruments and Methods in Physics Research A, 582, 733, (2007)

\bibitem{campana13} Campana, R., Feroci, M.,  Del Monte, E., Mineo, T.,
Lund, N.,  Fraser, G.W.,
Background simulations for the Large Area Detector
Experimental Astronomy, 29, (2013)

\bibitem{cusumano10} Cusumano, G., La Parola, V., Segreto, A., 
Ferrigno, C., Maselli, A.,  Sbarufatti, B.,  Romano, P.,  Chincarini, G.,  Giommi, P.,  Masetti, N.,
 Moretti, A., Parisi, P., Tagliaferri, G.
The Palermo Swift-BAT hard X-ray catalogue. III. Results after 54 months of sky survey
 Astronomy and Astrophysics, 24,  64,  (2010)


\bibitem{dsuza96}
D'Souza, A., Pantano, C.G.
Surface layer formation due to leaching and heat treatment of alkali lead silicate glass
Physics and chemistry of glasses , 37, 79, (1996)   

\bibitem  {feroci2010} 
Feroci, M., Stella, L., Vacchi, A., Labanti, C., Rapisarda, M., Attin{a}\', P., Belloni, T., 
Campana, R., Campana, S., Costa, E., Del Monte, E., Donnarumma, I., Evangelista, Y., 
Israel, G. L., Muleri, F., Porta, P., Rashevsky, A., Zampa, G., Zampa, N., Baldazzi, G., 
Bertuccio, G., Bonvicini, V., Bozzo, E., Burderi, L., Corongiu, A., Covino, S., 
Dall'Osso, S., de Martino, D., di Cosimo, S., di Persio, G., di Salvo, T., Fuschino, F., 
Grassi, M., Lazzarotto, F., Malcovati, P., Marisaldi, M., Mastropietro, M., Mereghetti, S., 
Morelli, E., Orio, M., Pellizzoni, A., Pacciani, L., Papitto, A., Picolli, L., Possenti, A., 
Rubini, A., Soffitta, P., Turolla, R., Zampieri, L.,
LOFT: a large observatory for x-ray timing, 
Proceedings of the SPIE, 7732, 1, (2010)

\bibitem {fraser10}
Fraser, G.~W.,  Carpenter, J.~D.,  Rothery, D.~A.,  Pearson, J.~F.,  Martindale, A.,  
Huovelin, J., Treis, J.,  Anand, M.,  Anttila, M.,  Ashcroft, M.,  Benkoff, J.,  
Bland, P.,  Bowyer, A.,  Bradley, A.,  Bridges, J.,  Brown, C.,  Bulloch, C.,  
Bunce, E.~J.,  Christensen, U.,  Evans, M.,  Fairbend, R.,  Feasey, M.,  
Giannini, F.,  Hermann, S.,  Hesse, M.,  Hilchenbach, M.,  Jorden, T.,  Joy, K.,  
Kaipiainen, M.,  Kitchingman, I., Lechner, P.,  Lutz, G.,  Malkki, A.,  Muinonen, K.,  
N\"ar\"anen, J.,  Portin, P.,  Prydderch, M., Juan, J.~S.,  Sclater, E.,  Schyns, E.,  
Stevenson, T.~J.,  Str\"uder, L.,  Syrjasuo, M.,  Talboys, D.,  Thomas, P.,  
Whitford, C.,  Whitehead, S.,
The mercury imaging X-ray spectrometer (MIXCS) on bepicolombo,
Planetary and Space Science, 58, 79, (2010) 

\bibitem {fraser11}
Fraser, G.~W., Brunton, A.~N., Lees, J.~E., Pearson, J.~F., Willingale, R., Emberson, D.~L., 
Feller, W.~B., Stedman, M., Haycocks, J.,
Development of microchannel plate (MCP) x-ray optics,
Proceedings of the SPIE, 2011, 215, (2011)

\bibitem {fraser92}
Fraser, G.~W. , Lees, J.~E. , Pearson, J.~F. , Sims, M.~R.,  Roxburgh, K.,
X-ray focusing using microchannel plates,
Society of Photo-Optical Instrumentation Engineers (SPIE) Conference Series, 1546, 41, (1992)
 
\bibitem{gruber99} Gruber, D.~E., Matteson, J.~L., Peterson, L.~E., Jung, G.~V.,
The Spectrum of Diffuse Cosmic Hard X-Rays Measured with HEAO 1, 
Astrophysical Journal, 520, 124, (1999) 

\bibitem{kaaret92} 
Kaaret, P., Geissbuehler, P., Chen, A.,  Glavinas, E.,
 X-ray focusing using microchannel plates,
 Applied Optics (ISSN 0003-6935),  31, 7339, (1992)

\bibitem{Lees97} 
Lees, J.~E. and Pearson, J.~F.,
 A large area MCP detector for X-ray imaging,
Nuclear Instruments and Methods in Physics Research A,  384, 410, (1997)

 \bibitem{martin00} 
Martin, A.P.,
X-ray optics, 
University of Leicester, PhD thesis, (2000)

\bibitem {peacock81}
Peacock, A.,  Andresen, R.~D., Manzo, G., Taylor, B.~G., Villa, G., Re, S., Ives, J.~C., Kellock, S.
The gas scintillation proportional counter on EXOSAT,
Space Science Reviews, 30, 525, (1981) 
 
\bibitem{rashevsky02}
Rashevsky, A.,  Bonvicini, V.,  Burger, P.,  Piano, S.,  Piemonte, C.,  Vacchi, A.,
Large area silicon drift detector for the ALICE experiment,
Nuclear Instruments and Methods in Physics Research A, 485, 54, (2002)

\bibitem{revnivtsev08} 
Revnivtsev, M.,  Molkov, S., Sazonov, S.,
Large-scale variations of the cosmic X-ray background and the X-ray emissivity of the local Universe,
Astronomy and Astrophysics, 483, 425, (2008) 

 \bibitem {turner81}
Turner, M.~J.~L., Smith, A., Zimmermann, H.~U.,
The medium energy instrument on EXOSAT,
Space Science Reviews, 30, 513, (1981)
 
\bibitem {vacchi91}
Vacchi, A.,   Castoldi, A.,   Chinnici, S.,   Gatti, E.,  Longoni, A.,   Palma, F. 
Sampietro, M., Rehak, P., Kemmer, J.,
Performance of the UA6 large-area silicon drift chamber prototype,
Nuclear Instruments and Methods in Physics Research A, 306, 187, (1991)

\bibitem{wallace06}
Wallace, K., Collon, M., Bavdaz, M., Fairbend, R., S\'eguy, J., Krumrey, M.,
 Developments in glass micro pore optics for x-ray applications,
Proceedings of the SPIE, 6266, 62661A, (2006)

\bibitem{willingale98}
Willingale, R., Fraser, G.~W., Brunton, A.~N., Martin, A.~P.,
Hard X-ray imaging with microchannel plate optics",
Experimental Astronomy, 8, 281, (1998)

\bibitem{wiza79}
Wiza, J.~L., 
Microchannel plate detectors
Nuclear Instruments and Methods, 162, 587, (1979)
 
\bibitem{zane2012}
Zane, S., Walton, D., Kennedy, T., Feroci, M., Den Herder, J.-W., Ahangarianabhari, M., 
Argan, A., Azzarello, P., Baldazzi, G., Barret, D., Bertuccio, G., Bodin, P., Bozzo, E., 
Cadoux, F., Cais, P., Campana, R., Coker, J., Cros, A., Del Monte, E., De Rosa, Alessandra, 
Di Cosimo, S., Donnarumma, I., Evangelista, Y., Favre, Y., Feldman, C., Fraser, G., Fuschino, F., 
Grassi, M., Hailey, M. R., Hudec, R., Labanti, C., Macera, D., Malcovati, P., Marisaldi, M., 
Martindale, A., Mineo, T., Muleri, F., Nowak, M., Orlandini, M., Pacciani, L., Perinati, E., 
Petracek, V., Pohl, M., Rachevski, A., Smith, P., Santangelo, A., Seyler, J.-Y., Schmid, C., 
Soffitta, P., Suchy, S., Tenzer, C., Uttley, P., Vacchi, A., Zampa, G., Zampa, N., Wilms, J., 
Winter, B.,
A large area detector proposed for the Large Observatory for X-ray Timing (LOFT), 
Proceedings of the SPIE, 8443, 2, (2012)

\end{thebibliography}
\end{document}